\documentclass[11pt,a4paper]{article}
\usepackage{amsfonts,amssymb}
\usepackage{cite}
\usepackage[T2A]{fontenc}
\usepackage[cp1251]{inputenc}
\usepackage[english]{babel}
\usepackage{amsmath}%
\usepackage[unicode]{hyperref}

\newtheorem{definition}{Definition}
\newtheorem{proposition}{Proposition}

\title{\bf Invariant manifolds for the hyperbolic type\\ integrable equations and their applications}
\author{\bf I.T. Habibullin and   A.R. Khakimova}

\begin{document}
\maketitle

\abstract{We assign some kind of invariant manifolds to a given integrable PDE (its discrete or semi-discrete variant). First, we linearize the equation around its arbitrary solution $u$. Then we construct a differential (respectively, difference) equation compatible with the linearized equation for any choice of $u$. This equation defines a surface called a generalized invariant manifold. In a sense, the manifold generalizes the symmetry, which is also a solution to the linearized equation. In this paper, we concentrate on continuous and discrete models of hyperbolic type. It is known that such kind equations have two hierarchies of symmetries, corresponding to the characteristic directions. We have shown that properly chosen generalized invariant manifold allows one to construct recursion operators that generate these symmetries. It is surprising that both recursion operators are related to different parametrizations of the same invariant manifold. Therefore, knowing one of the recursion operators for the hyperbolic type integrable equation (having no pseudo-constants) we can immediately find the second one.
}

\normalsize

\section{Introduction}

In our articles \cite{HKP}-\cite{HKh_2} we proposed a direct method for searching Lax pairs and recursion operators for integrable models. The method is based on the construction of invariant manifolds for the linearization of the nonlinear integrable equation under consideration in a neighborhood of an arbitrary solution.

It is natural to expect that the invariant manifold for a linear differential equation is also linear. However, in \cite{HKh}-\cite{HKh_2} we noticed that invariant manifolds can also be nonlinear. More precisely, we use linear generalized invariant manifolds to construct recursion operators. We obtain Lax pairs from manifolds defined by nonlinear functions.
Note that the requirement for the existence of a higher order invariant manifold compatible with the linearized equation imposes a strict condition on the nonlinear equation itself. In fact, only integrable equations admit this property.

Discussion on the methods of constructing  Lax pairs and recursion operators can be found in the literature (see \cite{BobenkoSuris} -- \cite{ZakharovShabat79}).

We emphasize that our method uses a close idea to the well-known method of the Walkwist and Estabrook pseudo-potentials \cite{WahlquistEstabrook}, where both Lax equations are simultaneously sought. On the contrary, for the given integrable equation we take its linearization as one of the Lax equations and look for the second one which is not supposed to be linear. In fact, at the first stage we find a nonlinear Lax pair and then linearize it with the corresponding point transformation. Since we are looking for only one of the two equations, our method is efficient enough. The application of the algorithm is illustrated by examples of integrable equations \eqref{hyp_eq_SM} and \eqref{u11}.

It is well-known that integrable hyperbolic type PDE admits two hierarchies of symmetries corresponding to two characteristic directions. In the article we discuss the relation between  recursion operators describing the hierarchies. It is observed that for the equation \eqref{hyp_eq_SM} (as well as for the discrete equation \eqref{u11}) these two recursion operators are generated by different parametrizations  of one and the same generalized invariant manifold.

\section{Generalized invariant manifolds  for the hyperbolic type equations}

Our articles \cite {HKh}-\cite {HKh_2} are devoted mainly to integrable  evolutionary type models. Now we will focus on discrete and continuous equations of hyperbolic type. We note that in the hyperbolic case invariant manifolds are defined somewhat differently. We begin with the hyperbolic type PDE
\begin{equation}
u_{xy}=f(u,u_x,u_y).\label{hyp_eq}
\end{equation}

Let us first recall some important definitions. Concentrate on the equation 
\begin{equation}
g(u_k,u_{k-1},\dots,u,\bar{u}_1,\bar{u}_2,\dots,\bar{u}_m)=0\label{surf_hyp_dv}
\end{equation}
defining a surface in the space of the dynamical variables $u,u_1,\bar{u}_1,u_2,\bar{u}_2,\dots$ of the equation~\eqref{hyp_eq}, where the notations $u_j=\frac{\partial^ju}{\partial x^j}, \bar{u}_j=\frac{\partial^ju}{\partial y^j}$ are used. We consider differential consequences:
\begin{eqnarray}
&&g_1(u_{k+1},u_{k},\dots,u,\bar{u}_1,\bar{u}_2,\dots,\bar{u}_{m-1})=0,\label{surf_dv_g1}\\
&&g_2(u_{k-1},\dots,u,\bar{u}_1,\bar{u}_2,\dots,\bar{u}_m,\bar{u}_{m+1})=0\label{surf_dv_g2}
\end{eqnarray}
of the equation \eqref{surf_hyp_dv} obtained by applying the operators of the total differentiation $D_x$ and $D_y$ with respect to $x$ and respectively $y$: $g_1=D_xg$ and $g_2=D_yg$  and by excluding all of the mixed derivatives of the function $u$ due to the equation~\eqref{hyp_eq}. Also we excluded $\bar u_m$ from $D_xg$ and $u_k$ from $D_yg$ due to equation \eqref{surf_hyp_dv}.

The surface \eqref{surf_hyp_dv} is called an invariant manifold of the equation (\ref{hyp_eq}) if the following equation holds 
\begin{equation}
\left.D_xD_yg\right|_{\eqref{hyp_eq}-\eqref{surf_dv_g2}}=0.\label{invman_cond}
\end{equation}
Assume that neither of the functions $g_1,g_2$ vanishes identically. Then evidently the invariant surface is of a finite dimension.

In what follows we will use the linearization of the equation \eqref{hyp_eq} around its arbitrary solution $u=u(x,y)$:
\begin{equation}
U_{xy}=aU_x+bU_y+cU,\label{lin_hyp_eq}
\end{equation}
where $a=\frac{\partial f}{\partial u_x}, b=\frac{\partial f}{\partial u_y}, c=\frac{\partial f}{\partial u}.$ 

Let us consider a surface defined by the equation 
\begin{equation}
G(U_k,U_{k-1},\dots,U,\bar{U}_1,\bar{U}_2,\dots,\bar{U}_m;u,u_1,\dots,u_{k_1};\bar{u}_1,\bar{u}_2,\dots,\bar{u}_{m_1})=0\label{surf_lin_hyp_eq},\quad k,m\geq0
\end{equation}
where $U_s=\frac{\partial^s}{\partial x^s}U, \bar{U}_s=\frac{\partial^s}{\partial y^s}U$ are the dynamical variables of the equation \eqref{lin_hyp_eq}. Here the dynamical variables of the equation \eqref{hyp_eq} $u,u_1,\bar{u}_1,\dots$ are considered as parameters. Find the differential consequences of the equation \eqref{surf_lin_hyp_eq}
\begin{equation}
G_1(U_{k+1},U_{k},\dots,U,\bar{U}_1,\bar{U}_2,\dots,\bar{U}_{m-1};u,u_1,\dots,u_{k_1+1};\bar{u}_1,\bar{u}_2,\dots,\bar{u}_{m_1})=0\label{surf_lin_hyp_G1},\quad m>0,
\end{equation}
and
\begin{equation}
G_2(U_{k-1},U_{k-2},\dots,U,\bar{U}_1,\bar{U}_2,\dots,\bar{U}_{m+1};u,u_1,\dots,u_{k_1};\bar{u}_1,\bar{u}_2,\dots,\bar{u}_{m_1+1})=0\label{surf_lin_hyp_G2}, \quad k>0
\end{equation}
obtained by applying the operators $D_x,D_y$ to the function $G$ such that $G_1=D_xG$, $G_2=D_yG$ and then excluding all of the mixed derivatives of $u$ and $U$ by means of the equations \eqref{hyp_eq} and, respectively, \eqref{lin_hyp_eq}. 
We also excluded $\bar U_m$ from $D_xG$ and $U_k$ from $D_yG$ due to equation \eqref{surf_lin_hyp_eq}.

\begin{definition}
We call the surface obtained by \eqref{surf_lin_hyp_eq} a generalized invariant manifold of the equation \eqref{hyp_eq} if the condition 
\begin{equation}
\left.D_xD_yG\right|_{\eqref{hyp_eq},\eqref{lin_hyp_eq}-\eqref{surf_lin_hyp_G2}}=0\label{invman_lin_cond}
\end{equation}
is satisfied identically for all values of the variables $u,u_1,\bar{u}_1,\dots$.
\end{definition}
Number $k+m$ is the order of the manifold \eqref{surf_lin_hyp_eq}. Notice that if \eqref{surf_lin_hyp_eq} defines a generalized invariant manifold for the equation \eqref{hyp_eq} then  $G_1$ doesn't depend on $\bar u_{m_1}$ and similarly $G_2$ doesn't depend on $u_{k_1}$.

Our main idea is to apply the reasonings above in a converse way. We examen the question whether the three equations \eqref{lin_hyp_eq}-\eqref{surf_lin_hyp_G1} constitute the Lax triad for the equation \eqref{hyp_eq}? More precisely we expect that the following consistency condition 
\begin{equation}
\left.D_yG_1\right|_{\eqref{lin_hyp_eq},\eqref{surf_lin_hyp_G1}}=0\label{invman_lin_cond_G1}
\end{equation}
recovers equation \eqref{hyp_eq}. Below in the section 3 we show that such a viewpoint is meaningful and can be used to construct the Lax pairs for integrable equations of the form \eqref{hyp_eq}.

Actually equations (\ref{surf_lin_hyp_G1}), (\ref{surf_lin_hyp_G2}) provide alternative parametrizations of the generalized invariant manifold, defined by the equation \eqref{surf_lin_hyp_eq}. They are obtained by applying the operators $D_x$ and $D_y$ to \eqref{surf_lin_hyp_eq} and by some further elementary transformations. It is clear that by iterating this procedure we can find special kind of parametrizations which are given as ordinary differential equations for the function $U(x,y)$ of the following form 
\begin{equation}
G_3(U_{k+m},U_{k+m-1},\dots,U;u,u_1,u_2,\dots)=0\label{G3}
\end{equation}
and
\begin{equation}
G_4(\bar U_{k+m},\bar U_{k+m-1}\dots,U;u,\bar u_1, \bar u_2,\dots)=0.\label{G4}
\end{equation}
Notice that $G_3$ depends only on $U$ and $u$ and their derivatives with respect to $x$, meanwhile $G_4$ depends on $U$ and $u$ and their derivatives with respect to $y$. Transition from equation \eqref{G3} to \eqref{G4} is discussed in section 4.

In some cases the condition $D_xG=0$ (or condition $D_yG=0$) holds identically and therefore doesn't define any parametrization of the manifold. In fact this means that $G$ is the $x$-integral (or $y$-integral) for the linearized equation \eqref{lin_hyp_eq}. Therefore, due to the well-known theorem (see, the survey \cite{ZhiberSokolov}) equation \eqref{hyp_eq} also admits a nontrivial $x$-integral (respectively $y$-integral). In what follows we suppose that equation \eqref{hyp_eq} doesn't admit any non-trivial $x$- and $y$-integrals. This kind integrals are called also pseudo-constants.

The invariant manifolds of special kind (\ref{G3}), (\ref{G4}) are closely connected with the symmetries of the equation \eqref{hyp_eq}. 

\section{Invariant manifolds and symmetries}

In this section we establish an important relation between the invariant manifolds of the hyperbolic type PDE and its evolutionary type symmetries. 

Recall that an evolutionary type PDE of the form
\begin{equation}
u_{t}=g(u,u_1,u_2,...u_l).\label{xsymm}
\end{equation}
is called a symmetry of the equation \eqref{hyp_eq} on the direction of $x$ if the following condition 
\begin{equation}
\left.D_xD_yg-D_tf\right|_{\eqref{hyp_eq},\eqref{xsymm}}=0\label{symm-cond}
\end{equation}
is satisfied identically for all values of the dynamical variables $u,u_1,\bar u_1,\dots$. In a similar way the symmetry on the direction of $y$ is defined.

An ordinary differential equation 
\begin{equation}
s(u,u_1,u_2,...u_{n})=0\label{im-symm}
\end{equation}
defines an invariant manifold for the equation \eqref{xsymm} if the following condition is satisfied
\begin{equation}
\left.D_ts\right|_{\eqref{xsymm}, \eqref{im-symm}}=0.\label{im-symm-cond}
\end{equation}
Here $D_ts$ is evaluated by means of \eqref{xsymm} and all of the derivatives $u_n,u_{n+1},\dots $ are expressed in virtue of the equation \eqref{im-symm}.

In what follows we will use the linearization of equation \eqref{xsymm}
\begin{equation}
U_{t}=a_lU_l+a_{l-1}U_{l-1}+\cdots +a_0U\label{lin-xsymm}
\end{equation}
where $a_j=\frac{\partial g}{\partial u_j}$ for $j=1,...l$. Let us define an ordinary differential equation 
\begin{equation}
U_{k}=H(U_{k-1},U_{k-2}, \dots,U;u,u_1,u_2,\dots)\label{H}
\end{equation}
where $U=U(x,t)$ is an unknown function and the dynamical variables $u,u_1,u_2,\dots$ of equation \eqref{xsymm} are considered as parameters.

We say that equation \eqref{H} defines a generalized invariant manifold for equation \eqref{xsymm} if the equation 
\begin{equation}
\left.D_tH-D_x^kU_t\right|_{\eqref{xsymm}, \eqref{lin-xsymm}, \eqref{H}}=0\label{gim-symm-cond}
\end{equation}
holds identically for all values of the variables $U,U_1,\dots U_{k-1};u,u_1, u_2,\dots$.
The following assertion about the relation of generalized invariant manifolds of the equation \eqref{hyp_eq} and of its symmetry sounds plausible.

{\bf Conjecture 1.} Let equation \eqref{xsymm} be a symmetry of equation \eqref{hyp_eq}. Then \eqref{H} defines a generalized invariant manifold for \eqref{xsymm} if and only if it defines a generalized invariant manifold for the equation \eqref{hyp_eq}.

\section{Invariant manifolds and recursion operators for the hyperbolic type integrable PDE}

Let us consider the following integrable hyperbolic type equation found in \cite{SokolovMeshkov}
\begin{equation}
u_{xy}=\sqrt{1+u_x^2}\sin u.\label{hyp_eq_SM}
\end{equation} 
Below we concentrate on the properties of the generalized invariant manifolds for this equation. By definition they are compatible with the linearized equation
\begin{equation}
U_{xy}=\sqrt{1+u_x^2}(\cos u)U+\frac{u_x\sin u}{\sqrt{1+u_x^2}}U_x.\label{lin_hyp_eq_SM}
\end{equation}
Equation \eqref{hyp_eq_SM} admits two hierarchies of higher symmetries \cite{SokolovMeshkov} corresponding to the characteristic directions of $x$ and $y$. It can be shown that the linear invariant manifold makes a bridge between these hierarchies. More exactly, the recursion operators corresponding to the hierarchies are derived from two different parametrizations of one and the same linear invariant manifold. Let us discuss the scheme in more details. In \cite{HKP} the following statement has been proved.

\begin{proposition}\label{proposition31}
Equation:
\begin{equation}
U_{xxx}-\left(\frac{u_{xx}}{u_x}+\frac{2u_xu_{xx}}{1+u_x^2}\right)U_{xx}+\left(u_x^2-\frac{u_xu_{xxx}}{1+u_x^2}+\frac{3u_x^2u_{xx}^2}{(1+u_x^2)^2}\right)U_x=\lambda^{-1}\left(U_x-\frac{u_{xx}}{u_x}U\right)\label{lin_uxxx}
\end{equation}
defines a generalized invariant manifold for the equation \eqref{hyp_eq_SM}, where $\lambda$ is a parameter.
\end{proposition}

Let us apply the operator $u_xD_x^{-1}\frac{1}{u_x}$ to \eqref{lin_uxxx} and obtain: 
\begin{equation}
R_{(x)}U=\lambda^{-1}U\label{Rx}
\end{equation}
where the operator 
\begin{equation*}
R_{(x)}=D_x^2-\frac{2u_xu_{xx}}{1+u_x^2}D_x+u_xD_x^{-1}\left(\frac{u_{xxx}}{1+u_x^2}-\frac{u_xu_{xx}^2}{(1+u_x^2)^2}+u_x\right)D_x 
\end{equation*}
is the recursion operator for the equation \eqref{hyp_eq_SM} in the direction of $x$. By applying $R$ to the r.h.s. of the classical symmetry $u_{\tau_1}=u_x$ we obtain the higher symmetry of \eqref{hyp_eq_SM} (cf. \cite{SokolovMeshkov}, \cite{SvinolupovSokolov})
\begin{equation}
u_\tau=u_{xxx}-\frac{3u_xu_{xx}^2}{2(1+u_x^2)}+\frac{1}{2}u_x^3.\label{utaux}
\end{equation}
Thus we have the following representation for the recursion operator $R_{(x)}$:
\begin{equation}
R_{(x)}=L_1^{-1}L_2\label{RxL1m1L2}
\end{equation}
where $L_1$ and $L_2$ are the differential operators 
\begin{equation*}
L_1=u_xD_x\frac{1}{u_x},\qquad L_2=D_{x}^3-\left(\frac{u_{xx}}{u_x}+\frac{2u_xu_{xx}}{1+u_x^2}\right)D_{x}^2+\left(u_x^2-\frac{u_xu_{xxx}}{1+u_x^2}+\frac{3u_x^2u_{xx}^2}{(1+u_x^2)^2}\right)D_x
\end{equation*}
which allow one to rewrite the invariant manifold \eqref{lin_uxxx} in a short form: 
\begin{equation*}
L_2U-\lambda^{-1}L_1U=0.
\end{equation*}
We reduce consecutively the order of the derivatives of $U$ with respect to $x$ in the formula \eqref{lin_uxxx} by using the following consequences of the equation \eqref{lin_hyp_eq_SM}:
\begin{eqnarray*}
&&1. \quad {} (D_y-a)U_x=bU;\\
&&2. \quad {} (D_y-a)U_{xx}=(a_x+b)U_x+b_xU; \\
&&3. \quad {} (D_y-a)U_{xxx}=(2a_x+b)U_{xx}+(a_{xx}+2b_x)U_x+b_{xx}U;
\end{eqnarray*}
where $a=\frac{u_x\sin u}{\sqrt{1+u_x^2}}, \, b=\sqrt{1+u_x^2}\cos u$.
By applying the operator $K_1=\frac{u_x^3}{au_{xx}}(D_y-a)$ to both sides of \eqref{lin_uxxx} we get the equation
\begin{equation}
U_{xx}-\left(\cot u+\frac{u_{xx}}{1+u_x^2}\right)u_xU_x+\frac{u_x\sqrt{1+u_x^2}}{\lambda \sin u}U_y-\frac{(1+u_x^2)}{\lambda}U=0\label{lin_uxx}
\end{equation}
which is a new parametrization of the invariant manifold. Next we apply to the obtained equation the operator $K_2=\frac{\sin^2u}{u_xu_y}(D_y-a)$ to find another parametrization of the invariant manifold 
\begin{equation}
U_x+\frac{\sqrt{1+u_x^2}\sin u}{\lambda u_y}U_{yy}-\frac{\sqrt{1+u_x^2}\cos u}{\lambda}U_y-(\lambda+\sin^2u)\frac{\sqrt{1+u_x^2}\sin u}{\lambda u_y}U=0.\label{lin_uyy}
\end{equation}
Afterward we apply to the last equation the following operator $K_3=\frac{\lambda u_y}{\sqrt{1+u_x^2}\sin u}(D_y-a)$ and obtain the required parametrization of the invariant manifold 

\begin{equation}
U_{yyy}-\frac{u_{yy}}{u_y}U_{yy}+(u_y^2-\sin^2u)U_y+\left(\frac{u_{yy}}{u_y}\sin^2u-3u_y\sin u\cos u\right)U=\lambda \left(U_y-\frac{u_{yy}}{u_y}U\right) \label{lin_uyyy}
\end{equation}
which can be rewritten in the following form, convenient for deriving the recursion operator $R_{(y)}$ in the $y$ direction 
\begin{equation}
\overline{L}_2U=\lambda\overline{L}_1U\label{mn_y_L2L1}
\end{equation}
where 
\begin{equation*}
\overline{L}_1=u_yD_y\frac{1}{u_y}, \qquad \overline{L}_2=D_y^3-\frac{u_{yy}}{u_y}D_y^2+(u_y^2-\sin^2u)D_y+\left(\frac{u_{yy}}{u_y}\sin^2u-3u_y\sin u\cos u\right).
\end{equation*}
Then finally we find $R_{(y)}=\overline{L}^{-1}_1\overline{L}_2:$
\begin{equation}
R_{(y)}=D_y^2+u_y^2-\sin^2u-u_yD_y^{-1}(u_{yy}+\sin u\cos u).\label{Ry}
\end{equation}
By applying $R_{(y)}$ to the classical symmetry $u_\tau=u_y$ we find the higher symmetry of \eqref{hyp_eq_SM} (see also \cite{SokolovMeshkov}, \cite{SvinolupovSokolov})
\begin{equation}
u_t=u_{yyy}+\frac{1}{2}u_y^3-\frac{3}{2}u_y\sin^2u.\label{uty}
\end{equation}

\section{Application of the scheme for finding the Lax pair}

In this section we show how to use the concept of the generalized invariant manifold for constructing the Lax pairs for the hyperbolic type integrable equations.
In the paper \cite{HKP} the following proposition is proved.
\begin{proposition}\label{proposition32}
Equation of the form 
\begin{equation}
U_y-\frac{\lambda \cos u}{\sqrt{1+u_x^2}}U_x-\frac{\sin u}{\sqrt{1+u_x^2}}\sqrt{(\lambda+1)((1+u_x^2)U^2-\lambda U_x^2)+c(1+u_x^2)}=0\label{sq_uy}
\end{equation}
defines a generalized invariant manifold for the equation \eqref{hyp_eq_SM}.
The corresponding equation of the form \eqref{surf_lin_hyp_G1} is as follows
\begin{equation}
U_{xx}-\frac{u_xu_{xx}}{1+u_x^2}U_x-(1+u_x^2)\lambda^{-1}U+\lambda^{-1}u_x\sqrt{(\lambda+1)((1+u_x^2)U^2-\lambda U_x^2)+c(1+u_x^2)}=0.\label{sq_uxx}
\end{equation}
\end{proposition}
In fact, \eqref{sq_uxx} is obtained from \eqref{sq_uy} by applying $D_x$ and then by excluding $U_y$ due to \eqref{sq_uy}.
Proposition 2 is easily proved by checking the consistency condition of the equations \eqref{lin_hyp_eq_SM}, \eqref{sq_uy}, \eqref{sq_uxx}.
We discuss how a nonlinear manifold \eqref {sq_uy} is obtained. It is derived from the known linear invariant manifold \eqref{lin_uyy} by imposing an additional constraint that reduces its order. In fact, we look for the restriction of the form
\begin{equation}
U_y=F(U,U_x,u,u_x)\label{uy_F}
\end{equation}
consistent with the equation \eqref{lin_uyy} for all values of the dynamical variables $u,u_x,u_y,\dots$. It is convenient to write equation \eqref{lin_uyy} in the following form
\begin{equation}\label{Uyy}
U_{yy}=-\frac{\lambda u_y}{\sqrt{1+u_x^2}\sin u}U_x+\frac{u_y\cos u}{\sin u}U_y+(\lambda+\sin^2u)U.
\end{equation}

Then evidently function $F$ should satisfy the equation
\begin{equation}
\left.D_y(F)-U_{yy}\right|_{\eqref{hyp_eq_SM},\eqref{lin_hyp_eq_SM},\eqref{uy_F},\eqref{Uyy}}=0,\label{uy_F_cond}
\end{equation}
which is rewritten in the following enlarged form:
\begin{eqnarray*}
\left.F_{U_x}U_{xy}+F_{U}U_{y}+F_{u_x}u_{xy}+F_{u}u_{y}+\frac{\lambda u_y}{\sqrt{1+u_x^2}\sin u}U_x-\frac{u_y\cos u}{\sin u}U_y-(\lambda+\sin^2u)U\right|_{\eqref{hyp_eq_SM},\eqref{lin_hyp_eq_SM},\eqref{uy_F}}=0.
\end{eqnarray*}
In the last equation we replace the variables $u_{xy}$, $U_{xy}$ and $U_y$ by means of the equations \eqref{hyp_eq_SM}, \eqref{lin_hyp_eq_SM} and \eqref{uy_F}, respectively. After some elementary transformations we obtain
\begin{eqnarray}
&&-\frac{u_y}{\sin u}\left(F(U,U_x,u,u_x) \sqrt{1+u_x^2}\cos u-F_u(U,U_x,u,u_x) \sqrt{1+u_x^2}\sin u-\lambda U_x\right)\nonumber\\
&&\qquad {} +F_{U_x}(U,U_x,u,u_x)((\cos u) U(1+u_x^2)+u_x(\sin u) U_x)+F_{u_x}(U,U_x,u,u_x)\sin u(1+u_x^2) \label{uy_F_cond_complete}\\
&&\qquad {} +F(U,U_x,u,u_x)F_{U}(U,U_x,u,u_x)\sqrt{1+u_x^2}-(\sin^2u+\lambda)\sqrt{1+u_x^2}U=0.\nonumber
\end{eqnarray}
Comparison of the coefficients at the independent variable $u_y$ in \eqref{uy_F_cond_complete} yields an ordinary differential equation for $F$:
\begin{equation*}
F(U,U_x,u,u_x) \sqrt{1+u_x^2}\cos u-F_u(U,U_x,u,u_x)\sqrt{1+u_x^2}\sin u -\lambda U_x=0\label{eq_at_uy}
\end{equation*}
which is easily solved
\begin{equation*}
F(U,U_x,u,u_x)=\frac{\lambda\cos u}{\sqrt{1+u_x^2}}U_x+F_1(U,U_x,u_x)\sin u.\label{solution_eq_at_uy}
\end{equation*}
By substituting the obtained specification of $F$ into \eqref{uy_F_cond_complete} we get an equation which splits down into the following two equations 
\begin{eqnarray}
&&1. \quad {} \frac{\sin u}{\cos u}\left((F_1(U,U_x,u_x)F_{1,U}(U,U_x,u_x)-U(\lambda+1))(u_x^2+1)^2\right.\nonumber\\
&&\qquad {} \left.+F_{1,u_x}(U,U_x,u_x)(1+u_x^2)^{5/2}+u_xU_xF_{1,U_x}(U,U_x,u_x)(1+u_x^2)^{3/2}\right)=0, \label{remain_eqs}\\
&&2. \quad {} (1+u_x^2)^{3/2}\left(F_{1,U_x}(U,U_x,u_x)U(1+u_x^2)+\lambda F_{1,U}(U,U_x,u_x)U_x\right)=0.\nonumber
\end{eqnarray}
The latter implies
\begin{equation*}
F_1(U,U_x,u_x)=F_2\left(u_x,\frac{(1+u_x^2)U^2-\lambda U_x^2}{1+u_x^2}\right).
\end{equation*}
We replace $F_1$ in the first equation in \eqref{remain_eqs} due to the obtained formula where we use notation $\theta=U^2-\frac{\lambda}{1+u_x^2}U_x^2$. As a result we get
\begin{equation*}
\left(2F_{2,\theta}(u_x,\theta)F_2(u_x,\theta)-\lambda-1\right)U+\sqrt{1+u_x^2}F_{2,u_x}(u_x,\theta)=0.
\end{equation*}
Since $U$ is an independent variable, here we have two equations which give $F_2(u_x,\theta)=F_2(\theta)$ and $\quad F_2(\theta)=\sqrt{(\lambda+1)\theta+c}$. Now we are ready to write down the final form of the searched function $F$. Obviously, \eqref{uy_F} reads as the equation
\begin{equation}\label{F_final}
U_y=\frac{\lambda \cos u}{\sqrt{1+u_x^2}}U_x+\frac{\sqrt{\lambda+1}\sin u}{\sqrt{1+u_x^2}}\sqrt{(1+u_x^2)U^2-\lambda U_x^2+c(1+u_x^2)/(\lambda+1)}
\end{equation}
which coincides with \eqref{sq_uy}. Evidently under the constraint \eqref{F_final} equation \eqref{lin_uyy} turns into \eqref{sq_uxx}.

Let us construct now a linear Lax pair for the equation \eqref{hyp_eq_SM} by using equations \eqref{sq_uy}, \eqref{sq_uxx}, \eqref{lin_hyp_eq_SM} where we put  $c=0$. To this end we introduce new variables $\varphi,\psi$ instead of $U, U_x$ by using the following quadratic forms
\begin{eqnarray}
&&U=\varphi^2+\psi^2,\label{Ufipsi}\\
&&U_x=\frac{2}{\sqrt{\lambda}}\sqrt{1+u_x^2}\varphi\psi.\label{Uxfipsi}
\end{eqnarray}
The consistency condition of \eqref{Ufipsi}, \eqref{Uxfipsi} gives rise to an equation 
\begin{equation}\label{fixpsix_1}
\varphi_x\varphi+\psi_x\psi-\frac{1}{\sqrt{\lambda}}\sqrt{1+u_x^2}\varphi\psi=0.
\end{equation}
Similary the consistency of \eqref{Uxfipsi} and \eqref{sq_uxx} with $c=0$ yields
\begin{equation}\label{fixpsix_2}
\varphi_x\psi+\psi_x\varphi+\frac{1}{2\sqrt{\lambda}}\left(u_x\sqrt{\lambda+1}\sqrt{(\varphi-\psi)^2(\varphi+\psi)^2}-\sqrt{u_x^2+1}(\varphi^2+\psi^2)\right)=0.
\end{equation}
Surprisingly the system of the equations \eqref{fixpsix_1}, \eqref{fixpsix_2} turned out to be linear 
\begin{gather}\label{fixpsix}
\left\{ \begin{array} {c} \varphi_x=\frac{1}{2\sqrt{\lambda}}\left(\sqrt{1+u_x^2}-\sqrt{\lambda+1}u_x\right)\psi,\\
\psi_x=\frac{1}{2\sqrt{\lambda}}\left(\sqrt{1+u_x^2}+\sqrt{\lambda+1}u_x\right)\varphi.
 \end{array}\right.
\end{gather}
It defines the $x$-part of the Lax pair. In order to obtain the $y$-part we apply the operator $D_y$ to both sides of the equations \eqref{Ufipsi}, \eqref{Uxfipsi} and simplify due to the equations \eqref{lin_hyp_eq_SM}, \eqref{F_final}. As a result, we get a linear equation again
\begin{gather}\label{fiypsiy}
\left\{ \begin{array} {c} \varphi_y=-\frac{1}{2}\sqrt{\lambda+1}\sin u\varphi+\frac{1}{2}\sqrt{\lambda}\cos u\psi,\\ 
\psi_y=\frac{1}{2}\sqrt{\lambda}\cos u\varphi+\frac{1}{2}\sqrt{\lambda+1}\sin u\psi.
 \end{array}\right.
\end{gather}
Equations \eqref{fixpsix}, \eqref{fiypsiy} constitute a Lax pair for the equation \eqref{hyp_eq_SM}.

By introducing a new spectral parameter $\xi$, due to the relation $\lambda=\frac{1}{4}(\xi-\xi^{-1})^2$ we arrive at the Lax pair depending rationally on $\xi$:
\begin{gather}
\Phi_x=\frac{1}{\xi-\xi^{-1}}\left( 
\begin{array} {cc} 0 & \sqrt{1+u_x^2}-\frac{u_x}{2}(\xi+\xi^{-1})\\ \sqrt{1+u_x^2}+\frac{u_x}{2}(\xi+\xi^{-1})& 0
\end{array}\right)\Phi,\label{fixpsix2} \\ \nonumber\\
\Phi_y=\frac{1}{4}\left( 
\begin{array} {cc} -(\xi+\xi^{-1})\sin u&(\xi-\xi^{-1})\cos u\\(\xi-\xi^{-1})\cos u& (\xi+\xi^{-1})\sin u
\end{array}\right)\Psi.\label{fiypsiy2}
\end{gather}
It has the singularities at points $\xi=\infty; 0; \pm1$.

\subsection{Comparison with the other Lax pair}

As it was observed in \cite{ZhiberSokolov} equation \eqref{hyp_eq_SM} is connected with the sine-Gordon equation 
\begin{equation}
v_{xy}=\sin v\label{hyp_sG}
\end{equation}
by the following differential substitution
\begin{equation}
v=u+ i arsinh (u_x), \quad i^2=-1\label{subs}
\end{equation}
where the function $y=arsinh(x)$ is defined from the equation $x=\sinh y$.

Therefore we can derive the Lax pair for the equation \eqref{hyp_eq_SM} by replacing $v, v_y$ due to \eqref{subs} in the Lax pair of the sine-Gordon equation \cite{Ablowitz}
\begin{gather}\label{sG-Lax}
\Psi_x=\frac{1}{2\lambda}\left( 
\begin{array} {cc}  \cos v&\sin v \\ \sin v& -\cos v
\end{array}\right)\Psi,  \qquad
\Psi_y=\frac{1}{2}\left( 
\begin{array} {cc} \lambda & -v_y\\ v_y& -\lambda
\end{array}\right)\Psi.
\end{gather}
As a result we get 
\begin{gather}
\Psi_x=\frac{1}{2\lambda}\left( 
\begin{array} {cc} \sqrt{1+u_x^2} \cos u -iu_x\sin u&\sqrt{1+u_x^2} \sin u +iu_x\cos u \\ \sqrt{1+u_x^2} \sin u +iu_x\cos u & -\sqrt{1+u_x^2} \cos u +iu_x\sin u
\end{array}\right)\Psi,\label{SvSoc-Lax_x}\\
\Psi_y=\frac{1}{2}\left( 
\begin{array} {cc} \lambda & -u_y-i\sin u\\ u_y+i\sin u& -\lambda
\end{array}\right)\Psi.\label{SvSoc-Lax_y} 
\end{gather}
It is easily verified that the consistency condition of the system  \eqref{SvSoc-Lax_x}, \eqref{SvSoc-Lax_y} is equivalent to the equation \eqref{hyp_eq_SM}.

The Lax pairs \eqref{fixpsix2}, \eqref{fiypsiy2} and \eqref{SvSoc-Lax_x}, \eqref{SvSoc-Lax_y} are connected with one another by the following gauge transformation $\Psi=S\Phi$ where 
\begin{gather}
S=\left( 
\begin{array} {cc} \xi p+iq & \xi q+ip\\ \xi q-ip& -\xi p+iq
\end{array}\right)
\end{gather}
with $p=\cos \frac{u}{2}-\sin \frac{u}{2}$ and $q=\cos \frac{u}{2}+\sin \frac{u}{2}$. The spectral parameters $\xi$ and $\lambda$ are related by the following equation $\lambda=\frac{1}{2}(\xi-\xi^{-1})$.

\section{Generalized invariant manifolds for the quad equations}
The scheme applied in the previous section can be adopted to the discrete case as well. Consider a discrete equation of the form
\begin{equation}
u_{n+1,m+1}=f(u_{n+1,m},u_{n,m+1},u_{n,m})\label{discrete_eq}
\end{equation}
defined on a quadratic graph, such that the sought function depends on two integers $n$ and $m$. To any of such equation one can assign an invariant manifold by analogy with the case of hyperbolic type PDE. Below we use the standard set of the dynamical variables of the equation (\ref{discrete_eq}) consisting of the variables in the set $\{u_{n+i,m}\}_{-\infty}^\infty \bigcup \{u_{n,m+j}\}_{-\infty}^\infty$. 

Let us concentrate on a surface in the space of the dynamical variables defined by the following equation
\begin{equation}
g(u_{n+s,m},\dots,u_{n+1,m},u_{n,m},u_{n,m+1},\dots,u_{n,m+k})=0.\label{discrete_eq_g}
\end{equation}
For the sake of definiteness we assume that the integers $s$ and $k$ are  nonnegative and at least one of them is positive. By applying the shift operators $D_n, D_m$ acting due to the rules $D_ny(n,m)=y(n+1,m)$ and $D_my(n,m)=y(n,m+1)$ to the equation (\ref{discrete_eq_g}) we obtain two additional equations 
\begin{gather}
g_1(u_{n+s+1,m},\dots,u_{n+1,m},u_{n,m},u_{n,m+1},\dots,u_{n,m+k-1})=0,\label{discrete_eq_g_1}\\ 
g_2(u_{n+s-1,m},\dots,u_{n+1,m},u_{n,m},u_{n,m+1},\dots,u_{n,m+k+1})=0\label{discrete_eq_g_2}
\end{gather}
where $g_1=D_ng$, $g_2=D_mg$.
\begin{definition}\label{definition_4}
Equation \eqref{discrete_eq_g} defines an invariant manifold for \eqref{discrete_eq} if the condition
\begin{equation}
\left.D_nD_mg\right|_{\eqref{discrete_eq}-\eqref{discrete_eq_g_2}}=0\label{DnDm_g}
\end{equation}
is satisfied.
\end{definition}
Let us study now a different situation. We define an invariant manifold not for the equation  \eqref{discrete_eq} itself, but for its linearization 
\begin{equation}
U_{n+1,m+1}=AU_{n+1,m}+BU_{n,m+1}+CU_{n,m}\label{lin_discrete_eq}
\end{equation}
where the coefficients are evaluated as follows $A=\frac{\partial f}{\partial u_{n+1,m}}, \quad B=\frac{\partial f}{\partial u_{n,m+1}}, \quad C=\frac{\partial f}{\partial u_{n,m}}$.
The definition of the invariant manifold discussed above can also be applied to the linearized equation \eqref{lin_discrete_eq} as well. However there is a peculiarity here since the coefficients $A,B,C$ of the equation depend of the dynamical variables $u_{n+i,m}$, $u_{n,m+j}$ of the equation \eqref{discrete_eq}. Therefore the linearized equation \eqref{lin_discrete_eq} is actually a family of the equations labeled by $u_{n,m}, u_{n+1,m}, u_{n,m+1},\dots$. 

We assign to equations \eqref{discrete_eq}, \eqref{lin_discrete_eq} a discrete equation
\begin{equation}
G(U_{n+s,m},\dots,U_{n+1,m},U_{n,m},U_{n,m+1},\dots,U_{n,m+k}; u_{n,m},u_{n\pm 1,m},u_{n,m\pm 1},\dots)=0,\label{discrete_eq_G}
\end{equation}
which depends on $u_{n,m}$ and its shifts, considered  as some parameters, while $U_{n,m}$ is interpreted here as the sought function.

Define the consequences of \eqref{discrete_eq_G} of the form
\begin{gather}
G_1(U_{n+s+1,m},\dots,U_{n+1,m},U_{n,m},U_{n,m+1},\dots,U_{n,m+k-1}; u_{n,m},u_{n\pm 1,m},u_{n,m\pm 1},\dots)=0,\label{discrete_eq_G_1}\\ 
G_2(U_{n+s-1,m},\dots,U_{n+1,m},U_{n,m},U_{n,m+1},\dots,U_{n,m+k+1}; u_{n,m},u_{n\pm 1,m},u_{n,m\pm 1},\dots)=0.\label{discrete_eq_G_2}
\end{gather}
Here functions $G_1$ and $G_2$ are obtained by applying the shift operators $D_n$ and $D_m$ and then excluding the mixed shifts $u_{n+i,m+j}$, $U_{n+i,m+j}$ on virtue of the equations \eqref{discrete_eq} and \eqref{DnDm_g}. Also we excluded the variable $U_{n,m+k}$ from $G_1=D_nG$ and the variable $U_{n+s,m}$ from $G_2=D_mG$ due to equation \eqref{discrete_eq_G}. 

We say that equation \eqref{discrete_eq_G} defines a generalized invariant manifold for the equation \eqref{discrete_eq_G} if the following equation 
\begin{equation}
\left.D_nD_mG\right|_{\eqref{discrete_eq},\eqref{lin_discrete_eq};\eqref{discrete_eq_G}-\eqref{discrete_eq_G_2}}=0\label{DnDm_G}
\end{equation}
is satisfied identically.

We defined above two transformations, one of them converts equation of the form \eqref{discrete_eq_G} to equation \eqref{discrete_eq_G_1} and the other converts the same equation to \eqref{discrete_eq_G_2}. By iterating the first of these transformations we can derive a parametrization of the form
\begin{equation}
G_3(U_{n+s_1,m},U_{n+s_1-1,m},\dots,U_{n,m}; u_{n,m},u_{n+1,m},\dots,u_{n+s_2,m})=0\label{discrete_eq_G_3}
\end{equation}
where $G_3$ depends only on the variables $U_{n,m}$, $u_{n,m}$ and their shifts with respact to $n$. Similarly, by iterating the second transformation we find the parametrization 
\begin{equation}
G_4(U_{n,m+k_1},U_{n,m+k_1-1},\dots,U_{n,m}; u_{n,m},u_{n,m+1},\dots,u_{n,m+k_2})=0\label{discrete_eq_G_4}
\end{equation}
where $G_4$ depends only on the variables $U_{n,m}$, $u_{n,m}$ and their shifts with respact to $m$.

Below in \S7 we illustrate that parametrizations \eqref{discrete_eq_G_3} and \eqref{discrete_eq_G_4} are related to the recursion operators for the equation \eqref{discrete_eq}.
We notice that the generalized invariant manifold can effectively be used for constructing the Lax pair to the equation \eqref{discrete_eq}. 

Recall that for the integrable models of the form (\ref{discrete_eq}) satisfying the consistency around a cube condition the algorithms of constructing the Lax pairs have been proposed in \cite{BobenkoSuris}, \cite{NijhoffWalker}  (see also \cite{Xenitidis}).

\section{An example of evaluating the Lax pairs and recursion operators for the quad equations via invariant manifolds}

Let us illustrate the application of the method of generalized invariant manifolds for constructing the recursion operators and the Lax pairs in the discrete case with an example. As a touchstone we take the well known discrete version of the KdV equation (see \cite{HirotaTsujimoto}, \cite{NijhoffCapel}):
\begin{equation}
u_{n+1,m+1}=u_{n,m}+\left(\frac{1}{u_{n+1,m}-u_{n,m+1}}\right).\label{u11}
\end{equation}
Below we use the abbreviated notation as follows. We put $u_{i,j}$ instead of $u_{n+i,m+j}$ and rewrite \eqref{u11} as $u_{11}=u+\frac{1}{u_{10}-u_{01}}$. Then the linearization of \eqref{u11} found due to \eqref{lin_discrete_eq} takes the form 
\begin{equation}
U_{11}=U-\frac{1}{(u_{10}-u_{01})^2}(U_{10}-U_{01}).\label{U11lin}
\end{equation}
In the article \cite {HKP} the following assertion has been  proved:
\begin{proposition}
Equation
\begin{equation}
U_{30}-\left(\lambda(u_{30}-u_{10})^2-\frac{u_{30}-u_{10}}{u_{20}-u}\right)U_{20}-\left(1-\lambda(u_{20}-u)(u_{30}-u_{10})\right)U_{10}-\frac{u_{30}-u_{10}}{u_{20}-u}U=0\label{U30lin}
\end{equation}
defines a generalized invariant manifold for the equation \eqref{u11}. Here $\lambda$ is an arbitrary constant.
\end{proposition}
Let us show that equation \eqref{U30lin} allows one  to write down  immediately the recursion operator in the direction of $n$ for the equation \eqref{u11}.  At first we shift the argument $n$ in \eqref{U30lin} backward and then divide the obtained equation by $u_{20}-u$. As a result we arrive at the equation 
\begin{eqnarray}\label{oim_U30m1}
&&\frac{1}{u_{20}-u}U_{20}+\frac{1}{u_{10}-u_{-10}}U_{10}-\frac{1}{u_{20}-u}U-\frac{1}{u_{10}-u_{-10}}U_{-10}\nonumber\\
&& \qquad {} \quad {} \qquad {} =\lambda\left((u_{20}-u)U_{10}-(u_{10}-u_{-10})U\right).
\end{eqnarray}
Afterward we apply the operator $\frac{1}{u_{10}-u_{-10}}(D_n-1)^{-1}$ to \eqref{oim_U30m1} and represent the result in the following form
\begin{eqnarray}\label{Rnulu}
R_{(n)}U=\lambda U,
\end{eqnarray}
where the operator
\begin{eqnarray}
R_{(n)}=\frac{1}{(u_{10}-u_{-10})^2}\left(D_n+D_n^{-1}\right)+\frac{2}{(u_{10}-u_{-10})(u-u_{-20})}+\nonumber\\ \frac{2}{u_{10}-u_{-10}}(D_n-1)^{-1}\left(\frac{1}{u-u_{-20}}-\frac{1}{u_{20}-u}\right)\label{Rn}
\end{eqnarray}
is the recursion operator for the equation \eqref{u11} in the direction of $n$. Evidently equation $u_{\tau}=0$ defines a symmetry for the equation \eqref{u11}. We obviously have the relation $R_{(n)}(0)=\frac{c}{u_{10}-u_{-10}}$ which defines the well-known symmetry of \eqref{u11} 
\begin{equation}
u_t=\frac{1}{u_{10}-u_{-10}}.\label{utn}
\end{equation}
On the symmetries of the quad equation \eqref{u11} see \cite{Tongas}, \cite{svinin}.

Since equation \eqref{u11} is invariant under the transformation $n\leftrightarrow m$ we can easily find the recursion operator $R_{(m)}$ in the direction $m$. However our goal here is to illustrate how to derive $R_{(m)}$ from known $R_{(n)}$ by some manipulations which can be used also in general case of the quad equation.

We apply the operator $D_m+\frac{1}{(u_{30}-u_{21})^2}$ to \eqref{U30lin} and in the obtained equation we replace the mixed shifts of the variables $u$, $U$ due to the equations \eqref{u11}, \eqref{U11lin}. After simplifications we get
\begin{eqnarray}
&&U_{20}-\frac{(u_{20}-u)((u_{10}-u_{01})(u_{20}-u)\lambda-1)}{u_{10}-u_{01}}U_{10}\nonumber\\
&&\qquad {} \quad {} \qquad {} +\left((u_{10}-u_{01})(u_{20}-u)\lambda-1\right)U+\frac{(u_{20}-u)(\lambda-1)}{u_{10}-u_{01}}U_{01}=0.\label{U20lin}
\end{eqnarray}
Let us repeat the same manipulation once again, i.e., we apply the operator $D_m+\frac{1}{(u_{20}-u_{11})^2}$  to  \eqref{U20lin} and after simplification due to \eqref{u11}, \eqref{U11lin}, we obtain
\begin{eqnarray}
&&U_{10}-\frac{(\lambda-1)(u_{10}-u_{01})}{u_{02}-u}U_{02}-\left((u_{10}-u_{01})(u_{02}-u)\lambda-\lambda+1\right)U_{01}\nonumber\\
&&\qquad {} \quad {} \qquad {} -\frac{((u_{10}-u_{01})(u_{02}-u)\lambda-\lambda+1)(u_{10}-u_{01})}{u_{02}-u}U=0.\label{U10lin}
\end{eqnarray}
Finally, we apply the operator $D_m+\frac{1}{(u_{10}-u_{01})^2}$ to \eqref{U10lin} and replace the variables $u_{11}$ and $U_{11}$ by \eqref{u11} and \eqref{U11lin}. As a result we get
\begin{eqnarray}
&&U_{03}-\frac{(u_{03}-u_{01})(1-\lambda-\lambda(u_{03}-u_{01})(u_{02}-u))}{(\lambda-1)(u_{02}-u)}U_{02}\nonumber\\
&&\qquad {} \quad {} \qquad {} +\frac{(1-\lambda-\lambda(u_{03}-u_{01})(u_{02}-u))}{\lambda-1}U_{01}-\frac{u_{03}-u_{01}}{u_{02}-u}U=0.\label{U03lin}
\end{eqnarray}
Let us show that generalized invariant manifold \eqref{U03lin} allows one to construct the recursion operator for the equation \eqref{u11} in the direction $m$. Indeed by shifting the argument $m$ backward we bring \eqref{U03lin} to the form
\begin{eqnarray}
\left(D_m^2+\left((u_{02}-u)^2+\frac{u_{02}-u}{u_{01}-u_{0,-1}}\right)D_m-((u_{02}-u)(u_{01}-u_{0,-1})+1)-\frac{u_{02}-u}{u_{01}-u_{0,-1}}D_m^{-1}\right)U=\nonumber\\
-\frac{u_{02}-u}{\lambda-1}(D_m-1)(u_{01}-u_{0,-1})U.\label{U03Rm}
\end{eqnarray}
We apply the operator $\frac{1}{u_{01}-u_{0,-1}}(D_m-1)^{-1}\frac{1}{u-u_{02}}$ to \eqref{U03Rm} and then write it as follows 
\begin{eqnarray}
R_{(m)}U=\frac{2-\lambda}{\lambda-1}U,\label{RmUlambdaU}
\end{eqnarray}
where
\begin{eqnarray}
&&R_{(m)}=\frac{1}{(u_{01}-u_{0,-1})^2}(D_m+D_m^{-1})+\frac{2}{(u-u_{0,-2})(u_{01}-u_{0,-1})}\nonumber\\
&&\qquad {} \quad {} \qquad {} +\frac{2}{u_{01}-u_{0,-1}}(D_m-1)^{-1}\left(\frac{1}{u-u_{0,-2}}-\frac{1}{u_{02}-u}\right)\label{Rm}
\end{eqnarray}
is the required recursion operator of equation \eqref{u11} in the direction $m$.

We now construct the Lax pair of the equation \eqref{u11} by means of generalized invariant manifolds. In article \cite{HKh} was proved the following proposition.
\begin{proposition}
Equation \eqref{U30lin} admits the first integral, which allows one to reduce its order. The corresponding first integral has the form 
\begin{eqnarray} 
U_{20}-U-(u_{20}-u)^2\lambda U_{10}-(u_{20}-u)\sqrt{4\lambda U_{10}U+c}=0. \label{oim_ut}
\end{eqnarray}
\end{proposition}

We set $c=0$ in \eqref{oim_ut} and by excluding $U_{20}$ from \eqref{oim_ut}, \eqref{U20lin} we obtain
\begin{eqnarray} \label{oim_U01sq}
(1-\lambda)U_{01}-U_{10}-\lambda(u_{10}-u_{01})^2U-2(u_{10}-u_{01})\sqrt{\lambda U_{10}U}=0.
\end{eqnarray}
Next, from the linearized equation \eqref{U11lin} and equation \eqref{oim_U01sq} we find
\begin{equation}
U_{11}=\frac{\lambda}{(1-\lambda)(u_{10}-u_{01})^2}U_{10}+\frac{1}{(1-\lambda)}U+\frac{2\sqrt{\lambda U_{10}U}}{(u_{10}-u_{01})(1-\lambda)}.\label{lin_U11sq}
\end{equation}

Let us change the variables $U=\hat{\varphi}^2$ in equations \eqref{oim_ut}, \eqref{oim_U01sq} and \eqref{lin_U11sq}. After elementary transformations we obtain a system of linear equations
\begin{eqnarray}
&& \hat{\varphi}_{20}=\sqrt{\lambda}(u_{20}-u)\hat{\varphi}_{10}-\hat{\varphi}, \label{lax_hfi20}\\
&& \hat{\varphi}_{01}=\frac{1}{\sqrt{1-\lambda}}\left(\hat{\varphi}_{10}-\sqrt{\lambda}(u_{10}-u_{01})\hat{\varphi}\right), \label{lax_hfi01}\\
&& \hat{\varphi}_{11}=\frac{1}{\sqrt{1-\lambda}}\left(\frac{\sqrt{\lambda}}{u_{10}-u_{01}}\hat{\varphi}_{10}-\hat{\varphi}\right). \label{lax_hfi11}
\end{eqnarray}

We simplify the triple \eqref{lax_hfi20}-\eqref{lax_hfi11} by means of the replacement $\hat{\varphi}=(1-\lambda)^{-\frac{m}{2}}\varphi$ and get
\begin{eqnarray}
&& \varphi_{20}=\sqrt{\lambda}(u_{20}-u)\varphi_{10}-\varphi, \label{lax_fi20}\\
&& \varphi_{01}=\varphi_{10}-\sqrt{\lambda}(u_{10}-u_{01})\varphi, \label{lax_fi01}\\
&& \varphi_{11}=\frac{\sqrt{\lambda}}{u_{10}-u_{01}}\varphi_{10}-\varphi. \label{lax_fi11}
\end{eqnarray}
We introduce a new variable $\psi$ in order to eliminate the variable $u_ {20}$ from the equation \eqref{lax_fi20}
\begin{eqnarray}\label{zam_psi}
\psi=\varphi_{10}-\sqrt{\lambda}u_{10}\varphi.
\end{eqnarray}
We apply the operator $D_n$ to the last equation
\begin{eqnarray}\label{zam_psidn}
\psi_{10}=\varphi_{20}-\sqrt{\lambda}u_{20}\varphi_{10}
\end{eqnarray}
and in the obtained equation we replace the variables $\varphi_{10}$ and $\varphi_{20}$ by virtue of equations \eqref{zam_psi} and \eqref{lax_fi20}, respectively:
\begin{eqnarray}\label{laxpsi10}
\psi_{10}=-(\lambda u_{10}u+1)\varphi-\sqrt{\lambda}u\psi.
\end{eqnarray}
Due to \eqref{zam_psi} we can rewrite equation \eqref{lax_fi01} as follows
\begin{eqnarray}\label{laxfi01}
\varphi_{01}=\sqrt{\lambda}u_{01}\varphi+\psi.
\end{eqnarray}
Now we  apply the operator $D_m$ to both sides of \eqref{zam_psi} and simplify it due to equations \eqref{u11}, \eqref{lax_fi11}, \eqref{zam_psi} and \eqref{laxfi01}:
\begin{eqnarray}\label{laxpsi01}
\psi_{01}=\left(\lambda-\lambda uu_{01}-1\right)\varphi-\sqrt{\lambda}u\psi.
\end{eqnarray}
Thus, we have received two systems:
\begin{equation}\label{lax10}
\left \{
\begin{array}{l}
\varphi_{10}=\sqrt{\lambda}u_{10}\varphi+\psi,\\
\psi_{10}=-(\lambda u_{10}u+1)\varphi-\sqrt{\lambda}u\psi,
\end{array} 
\right.
\end{equation}
\begin{equation}\label{lax01}
\left \{
\begin{array}{l}
\varphi_{01}=\sqrt{\lambda}u_{01}\varphi+\psi,\\
\psi_{01}=\left(\lambda-\lambda uu_{01}-1\right)\varphi-\sqrt{\lambda}u\psi,
\end{array} 
\right.
\end{equation}
which form the Lax pair of equation \eqref{u11}. We show that the pair found reduces to the already known Lax pair (see \cite{HirotaTsujimoto}, \cite{NijhoffCapel}). We set $\varphi=(-1)^{n+m}\lambda^{\frac{n+m}{2}}\tilde{\varphi}$, $\psi=(-1)^{n+m}\lambda^{\frac{n+m+1}{2}}\tilde{\psi}$. Then the pair \eqref{lax10}, \eqref{lax01} is written in the required form
\begin{equation}\label{laxu11}
\left \{
\begin{array}{l}
\Phi_{10}=A\Phi,\\
\Phi_{01}=B\Phi,
\end{array} 
\right.
\end{equation}
where 
\begin{equation*} \label{AB}
\Phi=\left( \begin{array}{c} \tilde{\varphi}\\ \tilde{\psi}\end{array} \right), \,
A=\left( \begin{array}{cc} -u_{10}&-1\\ uu_{10}+\lambda^{-1}& u\end{array} \right), \,
B= \left( \begin{array}{cc} -u_{01}&-1\\ uu_{01}+\lambda^{-1}-1& u \end{array} \right).
\end{equation*}

\section*{Conclusions}

In the integrability theory the linearized equation plays a crucial role. For instance, both classical and higher symmetries for a nonlinear equation are solutions of the linearized equation. We define a generalized invariant manifold to a nonlinear integrable equation as the invariant manifold to its linearization. Appropriately chosen generalized invariant manifold generates effectively the recursion operator as well as the Lax pair for the given equation. In fact the recursion operator corresponds to a linear generalized invariant manifold. Integrable equations of the hyperbolic type admit two hierarchies of symmetries and hence two recursion operators. These two recursion operators correspond to one and the same linear generalized  invariant manifold. Inspired by this observation we can conjecture that hyperbolic type integrable equation  which doesn't have non-trivial integrals in both characteristic directions possesses the following property: if it admits a hierarchy of higher symmetries in one characteristic direction then it admits the hierarchy of higher symmetries in the other direction as well.

\end{document}